\documentclass[twocolumn,showpacs,titlepage,aps,prd,letter,
tightenlines,amsmath,byrevtex,nofootinbib,floatfix]{revtex4}

\usepackage{graphicx}

\begin{document}


\title{The Cepheid Galactic Internet } 

\author{John G. Learned$^{1}$}
\email{jgl@phys.hawaii.edu}
\author{R-P. Kudritzki$^{2}$}
\email{kud@ifa.hawaii.edu}
\author{Sandip Pakvasa$^{1}$}
\email{pakvasa@phys.hawaii.edu}
\author{A. Zee$^{3}$}
\email{zee@kitp.ucsb.edu}

\affiliation{
$^1$Department of Physics and Astronomy, University of Hawaii, 
2505 Correa Road, Honolulu, Hawaii 96822 U.S.A.\\
$^2$Institute for Astronomy, Astronomy, University of Hawaii, 
2680 Woodlawn Drive, Honolulu, Hawaii 96822 U.S.A.\\
$^3$Kavli Institute for Theoretical Physics, University of 
California, Santa Barbara, California 93106 U.S.A. \\
}

\date{\today}

\vglue 1.6cm
\begin{abstract}

We propose that a sufficiently advanced civilization may employ Cepheid variable 
stars as beacons to transmit all-call information throughout the galaxy and beyond. 
One can construct many scenarios wherein it would be desirable for such a 
civilization of star ticklers to transmit data to anyone else within viewing range.  
The beauty of employing Cepheids is that these stars can be seen from afar (we 
monitor them out through the Virgo cluster), and any developing technological 
society would seem to be likely to closely observe them as distance markers.  
Records exist of Cepheids for well over one hundred years. We propose that these 
(and other regularly variable types of stars) be searched for signs of phase 
modulation (in the regime of short pulse duration) and patterns, which could be 
indicative of intentional signaling.

\end{abstract}


\maketitle

\section{Star Ticklers and the Need for a Galactic Lighthouse}
\label{introduction}

Considerations of communications with other life in the universe, the search for 
extraterrestrial intelligence (SETI) within (and beyond) our galaxy, have generally 
focused upon the use of microwaves and lasers as the communications medium. Problems 
arise from the difficult hurdles of resolution, noise and requisite power, making 
communications beyond our nearest neighbor stars very difficult, even with prior 
intention on both ends of a possible link. Searching for short (nanosecond) optical 
pulses shows some promise, though again suffers from the ``needle-in-a-haystack" 
problem. Some authors \cite{learned_2008} \cite{silagadze_2008} \cite{learned_1994} 
have proposed the use of neutrinos for avoiding some of the noise limitations, but 
large transmitter powers and formidable technical developments are still needed.

The problems with photons or neutrinos are not only data rate, but the ephemeral 
nature of the signal, if one is to contact an emergent civilization, one would 
have to waste enormous energy transmitting over aeons.  A better scheme may be 
to leave artifacts for an emergent civilization to discover as it becomes 
capable. Recently, some authors have driven home the point that it is far more 
energetically practical for transmitting large amounts of data to place long 
lasting artifacts in stellar systems to which the ETI may wish communicate 
information (their history for example) as intelligent life matures and becomes 
capable of decoding this "Rosetta stone"\cite{rose_2004}.

That still leaves a role for timely communication: some ``heads-up" type of 
communication informing newly technical societies of the existence of other life, 
perhaps of rules of engagement, instructions for finding said artifacts, or 
instructions on how to use some advanced means of communications (e.g. via 
extra-dimensions, or wormholes, or some other means beyond our present physics 
understanding).  We cannot guess what the motivation might be, but given a number of 
credible possible motivations, let us speculate upon what sort of galactic 
lighthouse such a civilization might employ.  We have speculated elsewhere 
\cite{learned_2008} about the possibility of employing high energy neutrino beams 
directed specifically at a candidate stellar system, even a specific planet. This 
might be important if there are some security concerns by the transmitting ETI.  
Another possibility, if such concerns do not dominate, which we explore herein, is 
to construct some method to reach everyone in the galaxy (and beyond). In another 
context, the use of the cosmic microwave background to reach everyone in the 
universe was also considered \cite{hsu_2006} but as far as we know that is not 
within the capability of any inhabitants of the universe.

In the following, we propose that the well studied Cepheid variables might 
provide an easily and likely to be monitored transmitter, which would be seen by 
all societies undertaking serious astronomy.

\section{Cepheid Variables}
\label{neutrino energies}

Cepheid variable stars was first observed in 1595.  They were first recognized 
as having the marvelous property of having a relationship between period and 
luminosity by Henrietta Swan Leavitt \cite{johnson_2005} in 1908, permitting the 
establishment of a distance ladder on the galactic scale.  The nearest stars 
could be ranged via parallax. Using the Cepheid scale one could move outwards up 
to stars in galaxies 20 megaparsec distant, and these stars have played a 
crucial role in the determination of the Hubble constant.  Cepheids are 
generally bright stars with significant modulation and are easily observed.  We 
expect that any civilization undertaking astronomy would soon discover them. Nor 
are there a daunting number of these, there being only of order 500 such stars 
presently tallied in our galaxy, and relatively few that are excellent 
standards.

The general picture for the Cepheids of Type I is that of a giant yellow 
star of population I with mass between five and ten times that of our sun, 
and $10^3$ to $10^4$ times the solar luminosity.  A dozen or so of these 
stars are visible to the naked eye. The period of the brightness excursion 
ranges between 1 and 50 days, and is generally stable. The approximate 
period is given by the stellar dynamical time scale, which is just 
radius/velocity $=(R^3 / G M)^{\frac{1}{2}}$. The period-luminosty 
relationship \cite{feast_1997} is given by $M_v = -2.81 ~log(P) - (1.43\pm 
0.1)$. The oscillation is explained by the build-up of ionized Helium, 
with increasing opacity with temperature, followed by violent expansion 
and de-ionization, and subsequent infall starting the cycle over again.

\section{How to modulate Cepheid periods}
\label{concept}

So, Cepheids form a class of readily detectable objects likely to be monitored by 
any emergent society which is undertaking astronomy. But can one modulate them?

The answer seems to us to be positive, because of the following.  Any oscillator of 
this type, a "blocking oscillator" for example, depends upon some integration over 
time (as with a capacitor in an electrical analogy) followed by a non-linear 
breakdown or reset. The system undergoes a phase change, and of necessity in such 
systems there is a period of instability.  A classic example of this is in heart 
beating \cite{wikswo_1999}, which can be triggered early with phase leading pulses.  
Near the time of normal triggering there is increased sensitivity to energy input.

We know, due to extensive modeling, that the solar luminosity is proportional to 
the 21$^{st}$ power of the core temperature.  In the sun the opacity decreases 
with temperature, and hence the sun is stable against perturbations.  In 
Cepheids the opposite is true \cite{zhevakin_1963} \cite{christy_1966} 
\cite{freedman_2008}, so that small depositions of energy can lead to runaway 
conditions, exactly what occurs naturally as the star settles back down from the 
previous excursion.  Delivery of a relatively small amount of energy at the 
right time can thus trip the cycle early.  One would thus expect that an early 
pulse would have somewhat less amplitude than a normal cycle.

Neutrinos would seem to be the ideal delivery means for transporting a pulse of 
energy to the stellar core, both due to penetration ability and to speed. If one 
wanted to employ infalling material as the trigger, this disturbance would propagate 
at near the local speed of sound, and might well evaporate prior to reaching the 
core in any event. Neutrinos of about 1 TeV would have an attenuation length of 
about $10^6$ km at solar densities. We have not modeled Cepheids in order to 
determine the optimum neutrino energy, but it is irrelevant for the present 
disucssion: we leave it as an engineering problem for the star tickling 
civilizations out there.  The initial flavor mix of the neutrino beam makes little 
difference since oscillations will mix the flavors and in any event the 
cross-sections are flavor independent.

Most Cepheid studies have worked with some form of Fourier transform of the sampled 
optical brightness (stellar magnitude), often with observational gaps.  For a 
regularly recurring pulsation with stable underlying period, this is adequate, and 
gaps can easily be dealt with. A long term periodogram will reveal the true period 
($\tau$). With phase advances present for encoding information, the long term 
Fourier transform will reveal a slightly shorter period on average, as determined by 
the magnitude and relative density of induced phase advances (call such period 
decrease $\tau\epsilon$). A simplest scheme would be with only one value of phase 
advance, a unique $\epsilon$, as opposed to a continuously variable or a set of 
integer values for $\epsilon$. For maximum information encoding, on average every 
other pulse would be shortened (a balanced set of ones and zeros). The information 
data rate $R$ would be then just one bit per period, reduced by half the average 
phase time shift: $R = 1/(\tau(1-\epsilon/2))$. The data encoded in a given 
observational sample would then have a Fourier transform which would be spread about 
this mean $R$, with sidebands depending upon the sample window, data rate and such, 
but with a split peak.  Given natural phase fluctuations, observational imprecision, 
and data windowing, the result may be hard to interpret.  Indeed peculiar spectra 
are seen \cite{welch_1993}, some of which have surely to do with complicating 
companion stars and are not what we seek. Others have complex structures due to more 
complicated oscillation modes.

One standard method of Cepheid analysis involves Dworetsky's "string minimizatio 
method \cite{dworetsky_1983}. Typically, a relatively small number of randomly 
spaced observations of a Cepheid are made over a long span of time. One guesses 
a trial period $\tau$ and plots the observed magnitude as a function of the 
phase as measured from the trial period. The points are then connected, giving a 
"string" in phase and magnitude space. The trial period is adjusted to achieve a 
minimum in this string length, thereby yielding both a period and a light-curve 
estimate.  The points typically scatter about the mean envelope, and often 
points are discarded as outliers \cite{welch_1993} \cite{stetson_1996} in order 
to find a smooth light-curve. Clearly such data manipulations could hide some 
unsuspected phase modulation of the data. In recent Cepheid studies some 
observers\cite{tanvir_2005} determine the light curve by a process they call 
"template fitting". This seems to be a good start for disentangling possible 
variations in the period which might be of a regular nature, but not discerned 
by Fourier transform.

We propose that a good way to examine this data for possible phase modulation would 
be to plot the histogram of measured pulse periods, deduced cycle by cycle, not 
averaged. This probably will result in some Gaussian-like distribution.  If however 
the distribution is multiply peaked, in the most ideal case having two equal peaks 
with discernible separation, then we may suspect the possibility of intentional 
modulation.  Should this be found, then we can interpret the identifications with 
one peak or the other as ones and zeros as in a binary encoded message.  Thereafter 
the conclusive finding of an ETI signal would come about by identification of 
regularities which are hard to understand from any natural oscillator, such as 
involving repeated complex sequences, prime numbers, a limited ``alphabet" (e.g. 
something akin to the familiar ASCII code), or even an apparent raster-like 
arrangement.  At this point we cannot but fall back upon the supposition that "we 
will recognize unnatural stellar fluctuations when we see them".  In sum, we think 
the normal method of employing Fourier transforms (with windowing, gap filling, long 
term averaging, etc.) and other methods in common use to analyze the periodicity of 
Cepheids may cloud the discovery of intentional modulation, and that cycle-by-cycle 
periods need to be examined.

\begin{figure}[htbp]
\begin{center}
\includegraphics[width=0.5\textwidth]{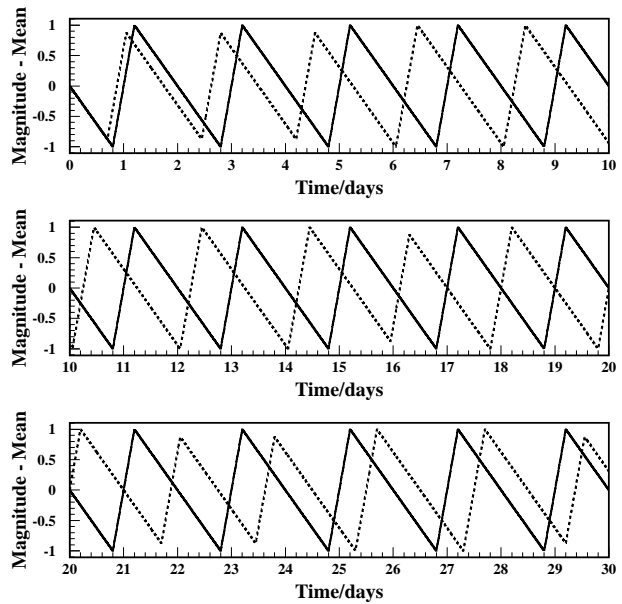}
\end{center}
\caption{Light curve of a simulated Cepheid vaiable. Ordinate is stellar 
magnitude relative to the mean, abscissa is time in days. The dashed 
curve represents an unmodulated (idealized) Cepheid with 2 day period and 2 
magnitude luminosity excursion, with expansion taking 0.4 days. The solid 
curve represents an arbitrarily modulated light curve with triggered 
phase advance of 0.1 day (0.05 cycle).  The units are, of course, 
arbitrary but representative of real data. The sharpness of the 
transitions which does not matter for the present discussions. } 
\label{fig:light_curve} 
\end{figure}

Of course with dense sampling of one or more observations per period, one can 
definitively disentangle the periods, pulse-by-pulse, and recover a "signal" 
should such be present. We have thought about the common situation of sparse 
observational sampling. We assume a simple saw toothed signal with longer and 
shorted periods, as illustrated in Figure \ref{fig:light_curve}. If one has a 
template for the waveform (light-curve, magnitude versus time), then with a 
single observation, one may extrapolate to (say) the descending zero crossing 
time, taking the mean magnitude as the zero point.  These zero crossing times 
then will be integer multiples of the fundamental Cepheid period, if no 
modulation is taking place.  If modulation is taking place, then the residuals 
will all be in clusters representing the number of cycles between observations.  
On the other hand if there is single phase modulation the first pulse will be 
doubled, the second pulse cluster will be tripled and so on.  This would clearly 
show evidence for binary cycle lengths.  Of course more complicated multiple 
level modulation might be employed by the ETI.  More difficult to discern would 
be the case of a continuous range of modulation.  Note also that in the example 
cited above we can have evidence for possible modulation, without being 
necessarily able to decode a string of ones and zeros: the phase plot may tell 
us that the modulation is very peculiar.

\begin{figure}[htbp]
\begin{center}
\includegraphics[width=0.5\textwidth]{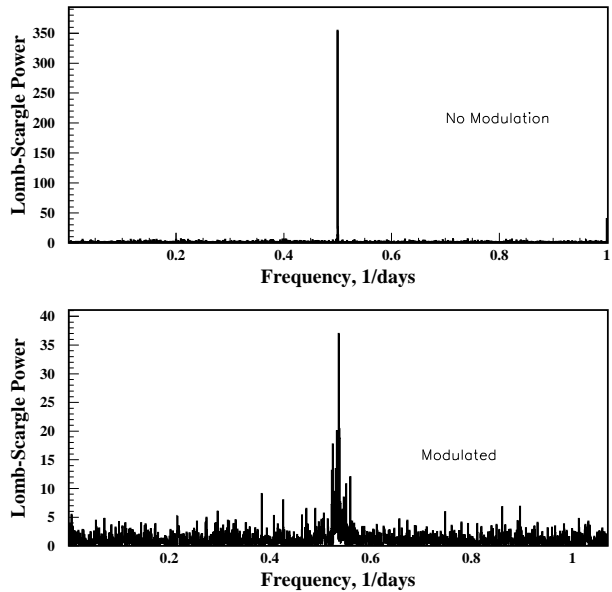}
\end{center}
\caption{Spectra of simulated observations of a regular periodic Cepheid 
variable and one with binary phase modulation. The ordinate is the 
Lomb-Scargle parameter, similar to chi squared; and abscissa is frequency.
Note that the more complicated structure of the modulated case is not
so obviously different from a noisey spectrum: one could not immediately 
discern that the latter case was not ``natural".} 
\label{fig:spectra}
\end{figure}

\begin{figure}[htbp]
\begin{center}
\includegraphics[width=0.5\textwidth]{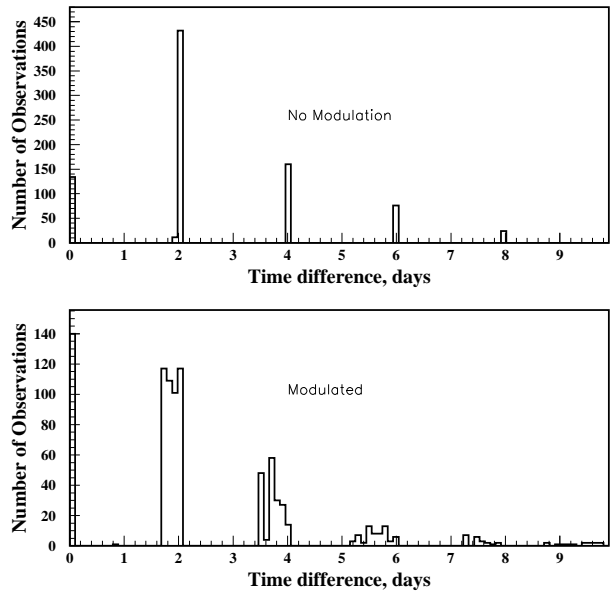}
\end{center}
\caption{Phase residuals of observations, when extrapolated to common
phase at period given by Lomb-Scargle peak. The unmodulated data shows 
peaks for obervations in the next cycle, one skipped cycle, two missed 
cycles, etc. The modulated case shows splitting of these cases
depending upon the combination of bits. This illustrates a possible 
means of detecting ``unnatural" phase variation without dense sampling.} 
\label{fig:phases} 
\end{figure}

In Figure \ref{fig:spectra} we show the Fourier transforms (or more properly the 
periodogram) of the unmodulated case (a delta function as one would expect) and a 
modulated spectrum.  In Figure \ref{fig:phases} we illustrate the phase residual 
technique for both unmodulated and modulated stars.  We take the period to be two 
days, with 1.8 days descending and 0.2 days expanding, and a total stellar magnitude 
range of 2.0.  We take the modulation to be 5\% in phase (or 0.1 days). The numbers 
are representative, but the actual choice has no particular meaning here, it simply 
illustrates the point that modulation will lead to peculiar phase residual plots.

This calculation is without question simplistic and meant to be merely illustrative. 
We imagine that experts in Cepheid analysis will be able to create more 
sophisticated algorithms. It may be noteworthy that there is a sort of inherent 
error correction coding in this type of modulation, in that the total phase offset 
with time tells one how many ones and zeroes have passed between observations.

\section{Energy Costs}
\label{energy}

Without detailed stellar modeling, we cannot make precise estimates of the amount of 
energy needed to trigger the Cepheid.  A brute force upper limit would be the 
addition of some reasonable fraction of the core energy production over a time scale 
of the speed of sound traversing the core.  Let us suppose that the pulse needs to 
be delivered in less than a second, and should be 10\% or more of the stellar 
output. If we take the Cepheid period to be 100,000 seconds, then the average power 
input in neutrinos should be about $10^{-6}$ of the stellar output.

However, this is probably an overestimate, since one may deliver the neutrino beam 
to one side of the core, triggering the conversion locally, as with a spark plug.  
We do not know the gain achievable by this method, only that it is surely large: the 
Cepheid is a huge signal amplifier.

There are also practical questions such as, how far away should the putative 
neutrino trigger generator be placed?  It cannot be too close to the star or it will 
(if consisting of any technology with which we are familiar) melt.  Such a device 
might sit out at a range of 100 AU, for example.  Perhaps it consists of an advanced 
``solar energy" power station capable of utilizing the radiation of the Cepheid 
itself, accumulated over a full cycle and then dumped into the neutrino trigger 
pulse. Located at a distance of 100 AU it would need to have receiving radius of 
about 0.2 AU.  It is premature for us to speculate too much about this, being enough 
to recognize that such unstable systems present the opportunity to serve as gigantic 
signal amplifiers, and we presume that the advanced civilization will be able to 
accomplish this, and speculate that they find the prospect worthwhile.

\section{Concluding Remarks}
\label{conclusion}

To summarize, in this note, we point out that that unstable stellar systems such 
as the Cepheids present an exciting opportunity to serve as gigantic signal 
amplifiers. We presume that a sufficiently advanced civilization will actually 
be able to tickle stars and we further speculate that they would find it 
worthwhile to use Cepheids as signaling devices. We propose that there may well 
be signatures of ETI communication available in data already recorded, and that 
a search of Cepheid (and perhaps other variable star, such as Lyrae) records may 
reveal an entre' into the galactic internet!

It may be a long shot, but should it be correct, the payoff would be immeasurable 
for humanity.  The beauty of this suggestion seems to be simply that the data 
already exists, and we need only look at the data in a new way.

\begin{acknowledgments}

We thank many colleagues for discussions of these ideas, in particular Freeman 
Dyson, Ralph Becker-Szendy, Walt Simmons, and Xerxes Tata. We would like to 
acknowledge support by the U.S.D.O.E. under grant DE-FG02-04ER41291 at the 
University of Hawaii, the N.S.F. under grant 04-56556 at the University of 
California at Santa Barbara.

\end{acknowledgments}


\end{document}